\documentclass[final,2p,times]{elsarticle}

\usepackage{amssymb}

\newcommand{\fe}{F_{esc}}
\newcommand{\px}{p_{max}}
\newcommand{\dK}{\rm{K}}
\newcommand{\ud}{{\rm d}}

\begin{document}

\begin{frontmatter}

\title{The contribution of supernova remnants to the galactic cosmic ray spectrum}

\author{D. Caprioli}
\ead{caprioli@arcetri.astro.it}

\author{E. Amato}
\ead{amato@arcetri.astro.it}

\author{P. Blasi}
\ead{blasi@arcetri.astro.it}

\address{INAF/Osservatorio Astrofisico di Arcetri, Largo E. Fermi, 5 - 50125 Firenze, Italy}
\address{Kavli Institute for Theoretical Physics, Kohn Hall, Santa Barbara (CA) 93106, United States}

\begin{abstract}
The supernova paradigm for the origin of galactic cosmic rays has been deeply affected by the development of the non-linear theory of particle acceleration at shock waves. Here we discuss the implications of applying such theory to the calculation of the spectrum of cosmic rays at Earth as accelerated in supernova remnants and propagating in the Galaxy. The spectrum is calculated taking into account the dynamical reaction of the accelerated particles on the shock, the generation of magnetic turbulence which enhances the scattering near the shock, and the dynamical reaction of the amplified field on the plasma. Most important, the spectrum of cosmic rays at Earth is calculated taking into account the flux of particles escaping from upstream during the Sedov-Taylor phase and the adiabatically decompressed particles confined in the expanding shell and escaping at later times. We show how the spectrum obtained in this way is well described by a power law in momentum with spectral index close to -4, despite the concave shape of the instantaneous spectra of accelerated particles. On the other hand we also show how the shape of the spectrum is sensible to details of the acceleration process and environment which are and will probably remain very poorly known. 
\end{abstract}


\end{frontmatter}

\section{Introduction}
\label{sec:intro}

In its original form \cite{GinzSiro}, the supernova remnant (SNR) paradigm for the origin of Galactic cosmic rays (CRs) is based on a purely energetic ground: if $\sim 10-20\%$ of the kinetic motion of the expanding shell of a supernova gets converted into accelerated particles, and one accounts for the energy dependent escape time from the Galaxy, SNRs can be the sources of the bulk of Galactic CRs. After the pioneering works on diffusive shock acceleration (DSA, \cite{krim77,bo78,bell78}), it became clear that this mechanism is the most promising acceleration process that can be responsible for energy conversion from bulk kinetic motion of a plasma to kinetic energy of charged particles. The DSA naturally leads to spectra of accelerated particles $N(E)\propto E^{-2}$ for strong shocks, not too dissimilar from the ones needed to describe data after accounting for the energy dependent escape time from the Galaxy, with a residence time that scales as $\tau_{esc}(E)\propto E^{-0.6}$. 

There are however two main concerns with this simple picture: first, the required acceleration efficiency is not so small that the dynamical reaction of the accelerated particles on the shock can be neglected. Second, if particle scattering is guaranteed by normal interstellar magnetic turbulence alone, the maximum energy of accelerated particles is exceedingly small and the mechanism cannot account for cosmic rays with energies up to the knee. It was soon understood that this second problem could be mitigated only by requiring CRs to generate the turbulence necessary for their scattering though streaming instability \cite{bell78,lagage}, a mechanism similar to that discussed by \cite{wentzel} in the context of CR propagation in the Galaxy. This latter point intrinsically makes the acceleration process even more non-linear.

The modern non-linear (NL) theory of DSA allows us to describe particle acceleration at SNR shocks by taking into account 1) the dynamical reaction of the accelerated particles on the system, 2) the magnetic field amplification due to streaming instability, and 3) the dynamical reaction of the amplified magnetic field on the plasma. These effects are interconnected in a rather complex way, so that reaching the knee and having enough energy channelled into CRs are no longer two independent problems. The situation is in fact even more complex given that the evolution of the SNR in time depends on the environment.

A generic prediction of NLDSA is that the spectra of accelerated particles are no longer power laws but rather concave spectra. In the case of extremely modified shocks, the asymptotic shape of the spectrum for $E\gg 1$ GeV is $N(E)\propto E^{-1.2}$ (see e.g. \cite{je91,maldrury} for reviews on CR modified shocks) to be compared with the standard $E^{-2}$ spectrum usually associated to DSA. Instead of clarifying the situation, this bit of information made it more puzzling in that so flat spectra are hard to reconcile with the CR spectrum observed at Earth. 

In this paper we show how the application of NLDSA to SNRs leads to time-integrated spectra that are very close to power laws at energies below 10-100 TeV where most measurements of CR spectra are performed with high statistical significance. The crucial piece of physics to connect the acceleration process inside the sources to the spectrum observed at Earth is the escape flux: during the Sedov-Taylor phase of the evolution (and to a lesser amount also during the ejecta dominated phase) particles can escape from a SNR in the form of a spectrum peaked at the maximum momentum reached at any given time. Particles which do not escape are advected downstream, lose energy adiabatically and eventually escape at later times. We calculate the spectrum injected by a single SNR as the superposition of these two components under different assumptions. Indeed, the semi-analytical method adopted here not only allows for a complete treatment of NLDSA but, being computationally very cheap, also allows for a very wide scan of the parameter space and an unprecedented investigation of the poorly known pieces of physics that enter the problem.

For simplicity, here we focus on type I supernovae, which occur in the typical interstellar medium (ISM), while qualitative differences between these and type II supernovae are only discussed by considering expansion in a more rarefied, hotter ISM, but totally ignoring any spatial stratification of the circumstellar region, which might be characterized by winds, bubbles and other complex structures. 

We also limit our attention to the proton component, while the results on nuclei will be presented in an upcoming paper since the additional issues that appear in that case deserve a detailed discussion. The introduction of nuclei is a fundamental step in the field and is essential to explain the CR spectrum above the knee (see e.g. \cite{kascade} and \cite{bertaina} for a review).

\section{A back-of-the-envelope calculation of the escape flux from a SNR}\label{sec:benchmark}

The escape of cosmic rays from a SNR is a very difficult problem to tackle, both from the physical and mathematical point of view. One can envision that at some distance upstream of the shock the particle density (or current) gets sufficiently small that the particles are no longer able to generate the waves that may scatter them and lead to their return to the shock front. These are escaping particles. However, the location of this free escape boundary is not easily calculated from first principles and it is usually assumed to be a given fraction of the radius of the shock. An additional uncertainty is introduced by the fact that the shock dynamics changes in time. The evolution of a SNR is characterized by three phases: an ejecta dominated (ED) phase, in which the mass of material accumulated behind the blast wave is less than the mass of the ejecta; a Sedov-Taylor (ST) phase, that starts when the accumulated mass equals the mass of the ejecta; a radiative phase, when the shock dissipates energy through radiation. The SNR is expected to spend most of the time over which it is active as a CR factory in the ST phase, that typically starts $500-1000$ years after the initial explosion. 

The maximum momentum of accelerated particles during the ED phase is expected to increase with time \cite{lagage}. As discussed by \cite{escape}, this is due to the fact that magnetic field amplification is rather efficient and the shock speed stays almost constant during this stage. 
After the beginning of the ST phase, the shock velocity, and thus also the efficiency of magnetic field amplification, decrease with time: as a consequence, the maximum momentum, $p_{max}$, is expected to drop with time as well \cite{escape}. The process of particle escape from the upstream region becomes important. At any given time, the system is no longer able to confine the particles that were accelerated to the highest energies at earlier times, so these particles escape from the shock.
The instantaneous spectrum of the escaping particles at any given time is very much peaked around $p_{max}(t)$ \cite{freeescape}. This qualitative picture of particle escape is the one that we mimic by assuming the existence of a free escape boundary, but as stressed above, the escape phenomenon is likely to be much more complex than suggested by this simple picture. 

Before embarking in a detailed calculation including the non-linear effects, it is useful to illustrate the results of a back-of-the-envelope calculation, based on a test-particle approach. Let us consider a SNR shell with a time dependent radius $R_{sh}(t)$ expanding with velocity $V_{sh}(t)$ in a uniform medium with density $\rho_{0}$ and suppose that escaping particles have momentum $\px(t)$ and carry away a fraction $\fe$ of the bulk energy flux $\frac{1}{2}\rho_{0} V_{sh}(t)^{3}$. Let $N_{esc}(p)$ be the spectrum of cosmic rays inside the remnant, so that the energy contained in a range $\ud p$ around $p$ is 
\begin{equation}\label{eq:dep}
\ud \mathcal{E}(p)=4\pi p^{2} N_{esc}(p)pc~\ud p\, .
\end{equation} 
The energy carried away by particles escaping in a time interval $\ud t$ at time $t$ is 
\begin{equation}\label{eq:det}
\ud\mathcal{E}(t)=\fe(t)\frac{1}{2}\rho V_{sh}^{3}(t)4\pi R_{sh}(t)^{2}\ud t\,.
\end{equation}
In a general way we can write $R_{sh}(t)\propto t^{\nu}$, and thus $V_{sh}(t)\propto t^{\nu-1}$. 
Using these time-dependencies, and equating the two expression for $\ud\mathcal{E}$, one obtains
\begin{equation}\label{eq:Nesc} N_{esc}(p)\propto t^{5\nu-3}\fe(t)p^{-3}\frac{\ud t}{\ud p}\,.
\end{equation}
During the ST stage, $\px$ is determined by the finite size of the accelerator, therefore we require that the diffusion length $\lambda(p)$ at $\px(t)$ is a fraction $\chi$ of the SNR radius (free escape boundary):
\begin{equation}\label{eq:defchi}
\lambda(\px)\simeq D(\px)/V_{sh}=\chi R_{sh}\, .
\end{equation}
 
Assuming for the diffusion coefficient the generic form $D(p)\propto p^\alpha/\delta B^\gamma$ and, for a magnetic field scaling as $\delta B(t)\propto t^{-\mu}$, we obtain
\begin{equation}\label{eq:pt}
	p(t)^\alpha\propto R_{sh}(t)V_{sh}(t) \delta B(t)^\gamma\propto t^{2\nu-1-\gamma \mu}\, ,
	\end{equation}
which implies
\begin{equation}\label{eq:dpt}
	\frac{\ud t}{\ud p}\propto \frac{t}{p}\, .
\end{equation}
Substituting Eq.~\ref{eq:dpt} into Eq.~\ref{eq:Nesc} one obtains:
\begin{equation}\label{eq:p4}
N_{esc}(p)\propto p^{-4}t^{5\nu-2} \fe(t);\quad t=t(p). 
\end{equation}
This relation illustrates a striking result: if the fraction of the bulk energy going into escaping particles  is roughly constant in time, and if the SNR evolution during the ST stage is adiabatic and self-similar (i.e.\ $\nu=2/5$), the global spectrum of particles escaping the system from the upstream boundary is exactly $p^{-4}$.
This means that the diffuse CR spectrum, usually explained by invoking the quasi-universal slope predicted by Fermi's mechanism at strong shocks, may be as well due the equally general evolution of a SNR during the ST stage. 

Possible corrections to Eq.~\ref{eq:p4} might lead to a slightly different spectrum for the escape flux. 
For instance, if the SNR evolution were not perfectly adiabatic, e.g.\ as a consequence of the energy carried away by escaping particles ($\nu\leq 2/5$) or if $\fe$ decreased with time (corresponding to a reduction of the shock modification), the spectrum of the escaping particles could be as flat as $\sim p^{-3.5}$. Reasonable modifications to the basic prediction for the escaping particle spectrum generally lead to spectra that are somewhat flatter than $p^{-4}$.

As we stressed above, the phenomenon of particle escape from the accelerator is very complex: for instance, in general the maximum momentum reached by particles at late stages of the SNR evolution is still high enough that there is a reservoir of CRs downstream that lose energy adiabatically during the expansion of the remnant and are eventually free to escape only when the shock dies out. The spectrum observed at Earth is made of the sum of these two components released at different times and with very different spectra.

\section{NLDSA at SNR shocks}

In this work we adopt the semi-analytical formalism for NLDSA developed by \cite{freeescape}, which represents the generalization of the work of \cite{ab05,ab06} to the case in which there is a free escape boundary at some position upstream of the shock. This calculation allows us to describe particle acceleration at a plane non relativistic shock in the assumption of quasi-stationarity and taking into account conservation of mass, momentum and energy, including the dynamical reaction of cosmic rays and amplified magnetic fields on the shock. The calculation makes use of the injection recipes discussed in \citep{bgv05}. 

In terms of mechanisms for magnetic field amplification, we only consider the (standard) resonant streaming instability, and the dynamical reaction of the amplified field on the plasma is taken into account as discussed in \cite{jumpkin}.
The assumption that only resonantly produced modes are excited in the upstream plasma is clearly rather restrictive, especially in the light of the recent results such as those by \cite{bell04} which suggest that non-resonant modes might grow faster and lead to more efficient magnetic field amplification at least during the early stages of the SNR evolution \cite{ab09}. On the other hand, such modes are typically produced at wavelengths which are much shorter than the gyration radius of the particles and can hardly be responsible for efficient scattering of particles at the highest energies, unless very rapid inverse cascading takes place. 

Possible damping of the magnetic field is also phenomenologically taken into account in a way that allows us to reproduce the results of \cite{pz03}: the damping efficiency is parametrized as
\begin{equation}\label{eq:damp}
	\zeta(t)=1-\exp\left[-\frac{V_{sh}(t)}{V_{damp}}\right]
\end{equation}
where $\zeta(t)$ is the ratio between the damping and growth rates. The results we present in the following are obtained with $V_{damp}$=2000 km/s, but we checked that varying  $V_{damp}$ varying between 500 and 5000 km/s leaves the results basically unchanged. The energy associated with damped magnetic turbulence is assumed to go into thermal energy of the background plasma (turbulent heating) as described in \cite{jumpkin} and references therein.

As already mentioned, from the point of view of the environment, we focus on SNRs in an ISM with spatially constant density. The circumstellar environment of type II/Ib,c SNe may be very complicated, depending on the details of the pre-SN stages (e.g.\ the production of Wolf-Rayet and Red Supergiant winds). We do not investigate here this possibly very complex structure and we qualitatively discuss the difference between type I and type II SNe by simply assuming a high density cold gas for the former and a rarefied warmer gas for the latter, just to illustrate the effects of these assumptions on the time integrated CR spectrum from a single remnant.

\begin{figure}
\begin{center}
\includegraphics[width=0.8\textwidth]{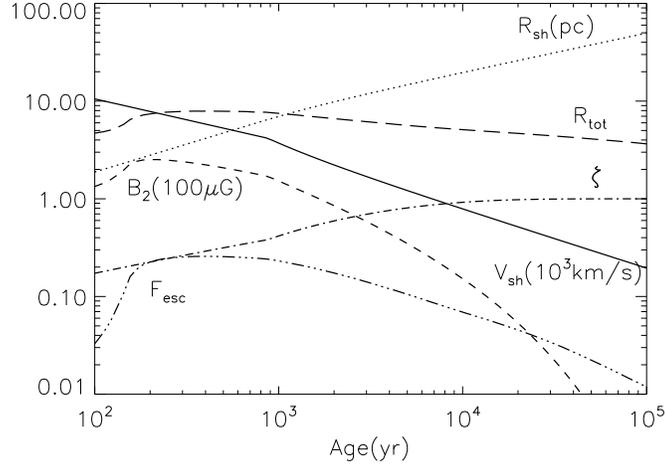}
\caption{Time dependence of the shock radius, $R_{sh}$ (in units of pc), shock velocity, $V_{sh}$ (in units of $10^3$ km/s), downstream magnetic field, $B_2$ (in units of $100\mu$G), flux through the free escape boundary placed at $x_{0}=R_{sh}$, $F_{esc}$ (in units of $\rho_0 V_{sh}^3/2$), total compression factor, $R_{tot}$, and damping parameter, $\zeta$ (see Eq.~\ref{eq:damp}). The curves refer to a SNR in a medium with magnetic field $B_{0}=5\mu$G, temperature $T_0=10^{5}\dK$, density $n_0=0.1 \rm{cm}^{-3}$, and injection parameter $\xi_{inj}=3.9$.\label{fig:hydro}} 
\end{center}
\end{figure}

The evolution of the forward shock position and velocity is taken as adiabatic and is described according to the analytical approach of \cite{TMK99} (table 7),  with $E_{SN}=10^{51}$erg for the SN energy and $M_{ej}=1.4M_\odot$ for the mass of the ejecta. In the case of modified shocks, this kind of solution is expected to only hold approximately, since escaping particles may in principle carry away a non-negligible amount of bulk energy, making the shock behave as partially radiative.

The evolution of the remnant is followed until its age is $\sim 10^{5}$ yr: for standard values of the parameters at this time $\px$  has dropped to values in the range 1-10 GeV/c. At each time-step the quasi-stationary solution for the shock dynamics and the instantaneous spectrum of accelerated particles is calculated. The calculation also returns the escape flux from $x=x_{0}$, the free escape boundary far upstream \cite{freeescape}. 
The flux of CRs contributed at Earth by a single remnant is the result of the integration over time of the instantaneous escape flux plus the spectrum of particles advected downstream and escaping at later times. Treatment of this latter part is especially problematic and requires some discussion. If diffusion is neglected behind the shock, in principle, particles that are advected downstream sit within a fluid element in which the strength of the magnetic field, in the absence of damping, is just the result of adiabatic decompression of the field just behind the shock at the time when these particles were accelerated. In this case some fraction of particles, even at the highest energies, may remain confined downstream and lose energy in the expansion of the shell. The escape of these accumulated particles will be possible only at very late times, after the shock has dissipated away. It is important to realize that in this scenario, due to adiabatic losses, none of the advected particles can actually escape at the knee energy. 

In order to describe adiabatic losses, we assume that the post-shock pressure, dominated by the sum of gas+CR pressure, is nearly uniform: a reasonable assumption, given that the fluid is subsonic. The downstream plasma pressure is proportional to the square of the shock Mach number, hence $\rho^{\gamma}(t)\propto p_{gas}(t)\propto V_{sh}^{2}(t)$. A relativistic particle with energy $E_{0}$ advected downstream at time $t_{0}$ will at a later time $t$ have an energy $E(t)=E_{0}/L(t_{0},t)$, with $L(t_{0},t)=\left[V_{sh}(t_{0})/V_{sh}(t)\right]^{\frac{2}{3\gamma}}$. It is possible to check {\it a posteriori} that choosing $\gamma=5/3$ or 4/3, respectively corresponding to a gas or CR dominated pressure, does not lead to major differences in the results.
Other authors have proposed that advected particles stop suffering adiabatic losses only when the pressure of the fluid element they sit with matches the ISM value \cite{bv83}. This recipe leads to very severe losses for the advected particles and when their spectrum is added to that contributed by the escaping particles, the result is very far from a power-law and incompatible with observations.

Either because of magnetic field damping or because of gradients in the magnetic field strength downstream (possibly induced by gradients in the accelerated particle pressure), it could well be that particles of a given maximum energy at a given time cannot be confined downstream at later times. In this case, at any time $t$ all particles with momentum $p\geq p_{esc}(t)$ must escape the system, where $p_{esc}(t)$ is defined so that the corresponding diffusion length in the instantaneous downstream magnetic field is 
$\lambda(p_{esc},B_{2})\sim x_0$. It is easy to show (and we will do it later) that $p_{esc}(t)\geq p_{max}(t)$ at any time.

These two recipes (escape at $p\sim p_{max}(t)$ and escape at $p_{esc}(t)$) lead to different integrated spectra from an individual SNR and unfortunately they are not the only two conceivable scenarios for particle escape. For instance large scale instabilities could break the structure of the forward shock in smaller size shocks that could allow some particle escape sideways. In this case it might make sense to assume that some fraction of the advected particles at any time may escape the system with their instantaneous spectrum. 

\begin{figure}
\begin{center}
\includegraphics[width=0.8\textwidth]{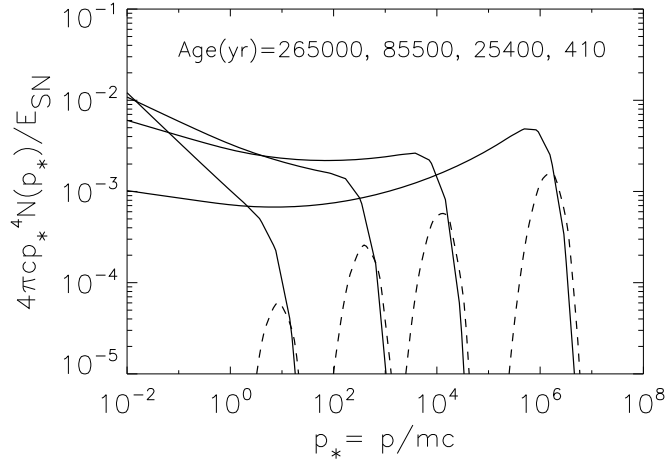}
\caption{Escape flux through the free escape boundary (dashed lines) and advected spectra (solid lines) at four different times for the same benchmark SNR used for Fig.~\ref{fig:hydro}} 
\label{fig:spectra}
\end{center}
\end{figure}

Given the importance of this physical phenomenon for establishing the spectrum of CRs observed at Earth, in the following we illustrate the time integrated spectra for the three escape scenarios outlined above. 

An attempt to calculating the cumulative injection spectrum of CRs has been made by \cite{pz05}: in their approach the shock modification was fixed {\it a priori} and constant throughout the whole SNR evolution, rather than being self-consistently calculated, and no dynamical feedback of the magnetic field was taken into account. However, the evolution of the shock modification is, in fact,  strictly connected with the acceleration efficiency and in turn with the normalization of the spectrum, so that only within a self-consistent non-linear approach it is possible to understand which are the SNR stages that contribute the most to the diffuse galactic CR spectrum.

For illustrative purposes, in Fig.~\ref{fig:hydro} we show the time dependence of the shock radius, $R_{sh}$, shock velocity, $V_{sh}$, downstream magnetic field, $B_2$, flux through the free escape boundary, $F_{esc}$, total compression factor, $R_{tot}$, and damping parameter, $\zeta$. 
The curves refer to a SNR expanding in a medium with background magnetic field $B_{0}=5\mu$G, temperature $T_0=10^{5}$K, density $n_0$=0.1 cm$^{-3}$; the injection parameter is fixed as $\xi_{inj}=3.9$ and $x_0=R_{sh}$.
In order to highlight the need for a non-linear treatment of DSA, it is worth noticing that at the beginning of the ST phase ($\sim 800\textrm{yr}$) the total compression ratio is $R_{tot}\sim 9$, corresponding to $\sim50\%$ of the bulk pressure channelled into CRs, and $F_{esc}\sim 20\%$.

We also show, in Fig.~\ref{fig:spectra}, the escape flux through the free escape boundary (dashed lines) and advected spectra (solid lines) at four different times (as specified on the figure) for the same benchmark SNR used for Fig.~\ref{fig:hydro}.

\subsection{Escape of particles around $p_{max}(t)$}

Here we focus on the escape recipe in which at any given time particles escape in a narrow region around $p_{max}$ as discussed in \cite{freeescape}, but most of the particles are advected downstream and stay there losing energy adiabatically.

\begin{figure}
\begin{center}
\includegraphics[width=0.49\textwidth]{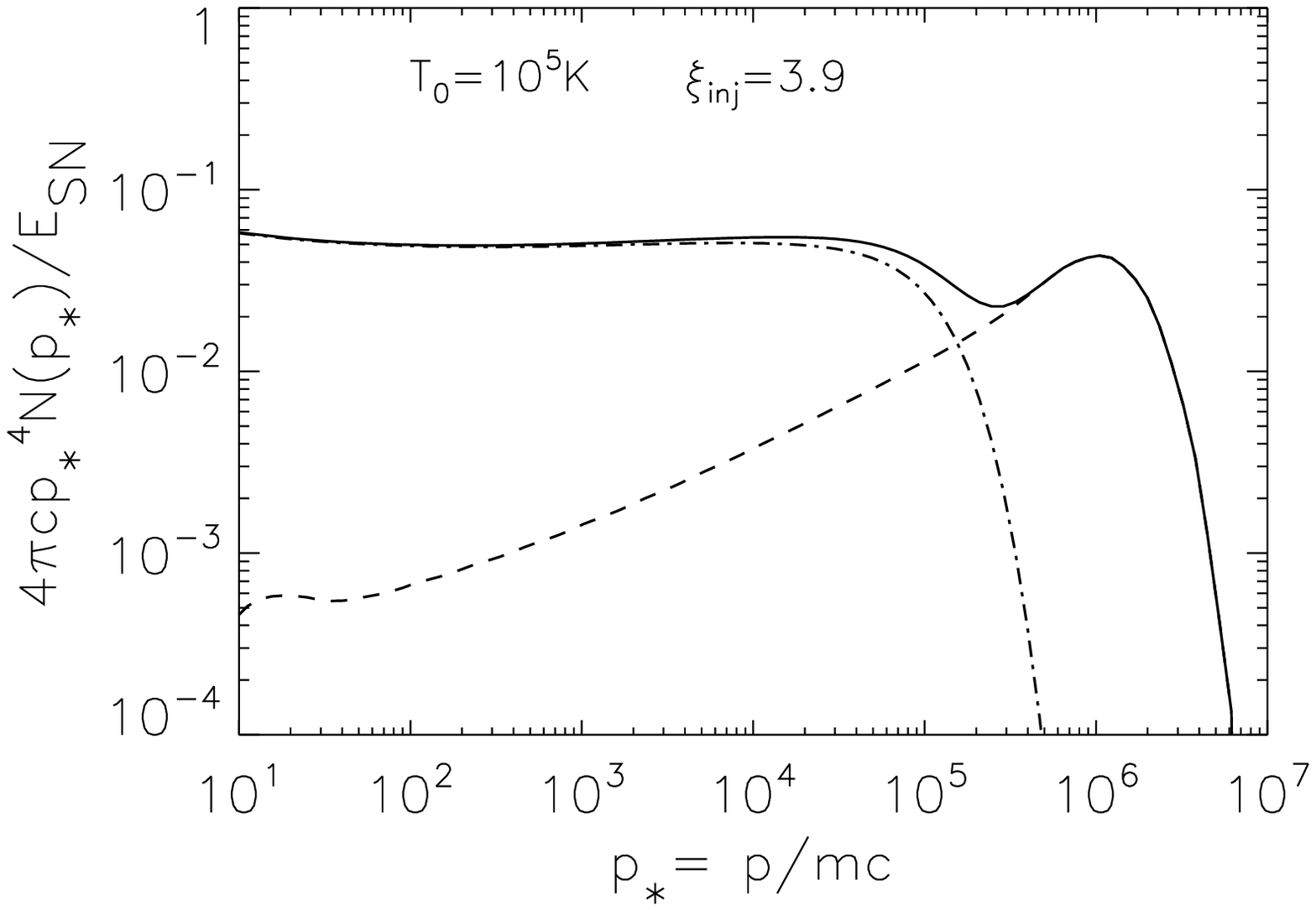}
\includegraphics[width=0.49\textwidth]{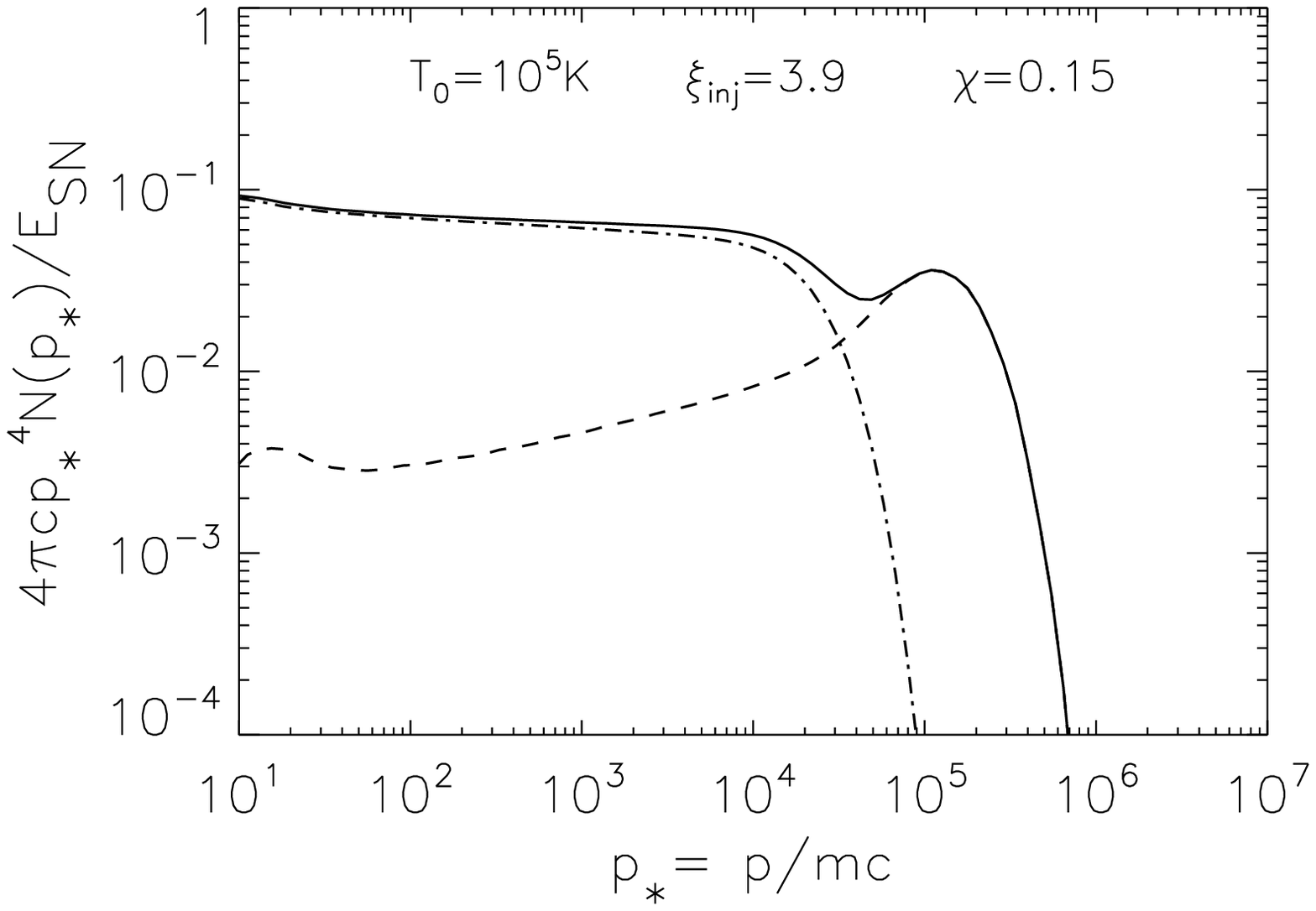}
\caption{CR spectrum injected in the ISM by a SNR expanding in a medium with density $n_0=0.1 \rm{cm}^{-3}$, temperature $T_{0}=10^{5}\dK$ and injection parameter $\xi_{inj}=3.9$. The dashed line is due to the escape of particles from upstream, the dash-dotted line is the spectrum of particles escaping at the end of the evolution. The solid line is the sum of the two. {\it Left}: $x_{0}=R_{sh}$. {\it Right}: $x_{0}=0.15 R_{sh}$.} 
\label{fig:T5xi39}
\end{center}
\end{figure}

In Fig.~\ref{fig:T5xi39} we illustrate the CR spectrum from our benchmark SNR, where we assume that the ISM has density $n_0=0.1 \rm{cm}^{-3}$ and temperature $T_{0}=10^{5}\dK$. The injection is assumed to correspond to $p_{inj}=3.9 p_{th,2}$ where $p_{th,2}$ is the momentum of thermal particles downstream of the shock (see \cite{bgv05}). The left panel refers to the case in which the free boundary condition is imposed at a distance from the shock $x_{0}=R_{sh}$, while the right panel refers to $x_{0}=0.15 R_{sh}$. The latter value of $x_0$ approximately corresponds to the position of the contact discontinuity at the beginning of the ST phase. This ratio, however, increases with time and becomes of order 1 before the beginning of the radiative phase.

The dashed lines represent the spectrum of particles that escape from the remnant towards upstream infinity at any given time. The peak at high energies corresponds to early times, when the maximum energy is the highest. At later times particles of lower and lower energy escape. The spectrum is somewhat flatter than $p^{-4}$ because the escape flux decreases with time, as discussed in \S \ref{sec:benchmark}. The dash-dotted lines represent the spectrum contributed by the particles trapped inside the remnant and escaping at the end of the evolution, after the effect of adiabatic losses. 
Here the SNR is assumed to die as a CR factory at an age of $\sim10^{5}$yr, namely when the amplified magnetic field has dropped below $\delta B/B_{0}<10^{-3}$ and thus $\px\sim$1-10 GeV/c.
The solid line, which is the sum of the two contributions, is very close to being the canonical power law $p^{-4}$. In this case, as in most cases we will show below, a dip is present in the spectrum. This dip is found at energies a factor of a few below the maximum one and marks the transition between energies at which the advected particles are the dominant contribution and energies where only escape at the early ST stage is important. The distance between the cutoff in the spectrum of advected particles and the peak at the highest energies provides an estimate of the strength of adiabatic energy losses. 

A few points are worth being noticed: 1) the accelerated particles reach the knee only if one chooses $x_{0}=R_{sh}$ (left panel), while for the more popular choice $x_{0}=0.15 R_{sh}$ (right panel), the maximum energy is appreciably lower. On the other hand this conclusion depends on the details of the magnetic field generation and scattering properties. We cannot exclude that more efficient magnetic field amplification on spatial scales which may be responsible for resonant scattering of particles at $p_{max}$ may change this conclusion. In this case however the general trend is to have somewhat flatter spectra, so that the naive expectation is that the time-integrated particle spectrum will resemble the one on the left panel.
2) the spectral concavity which is typical of NLDSA and that appears very clearly in the instantaneous particle spectra is almost completely washed out by the temporal evolution. In the case with $x_{0}=0.15 R_{sh}$, for example, the time convolution leads to spectra even slightly steeper than $p^{-4}$.

\begin{figure}
\begin{center}
\includegraphics[width=0.49\textwidth]{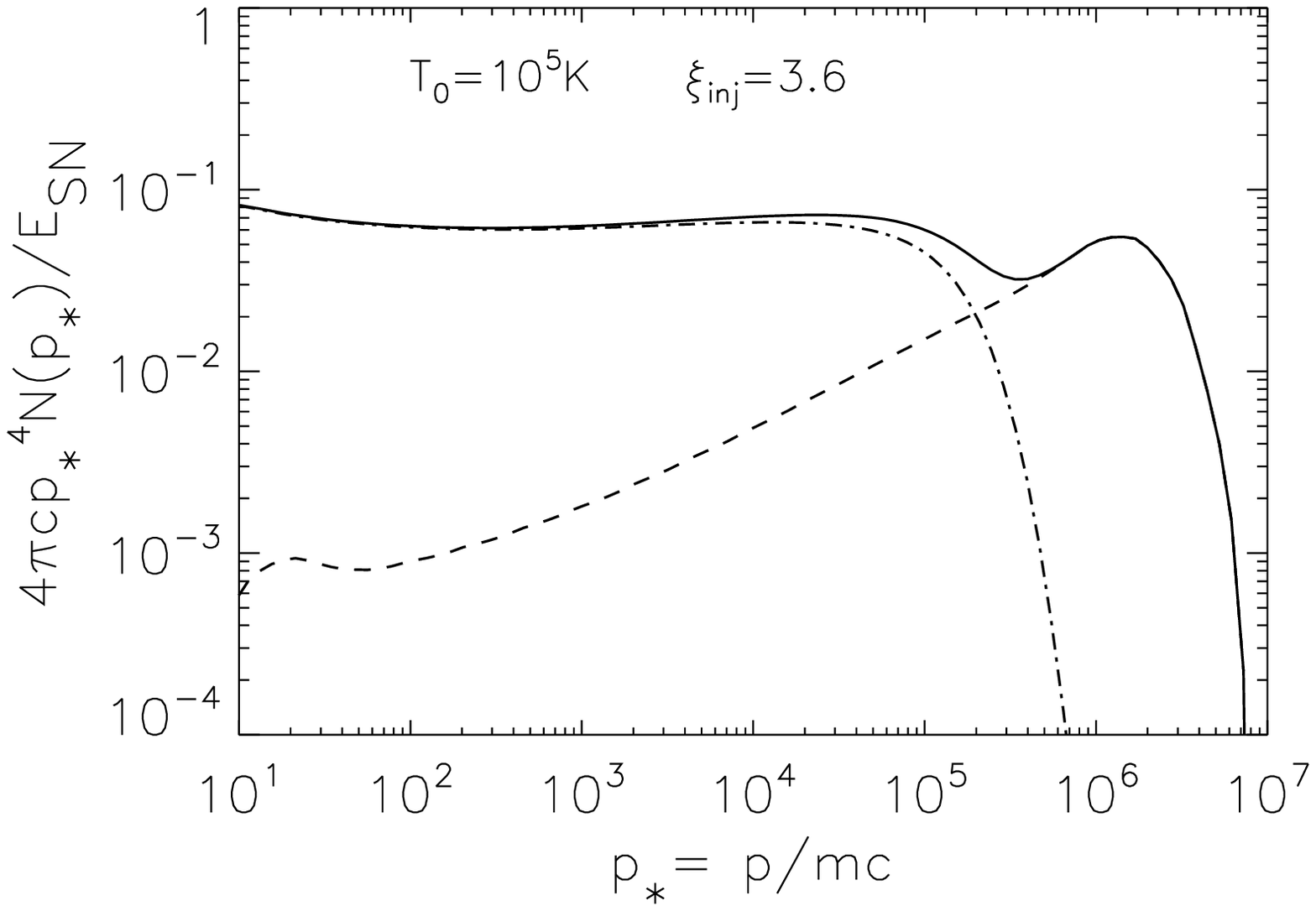}
\includegraphics[width=0.49\textwidth]{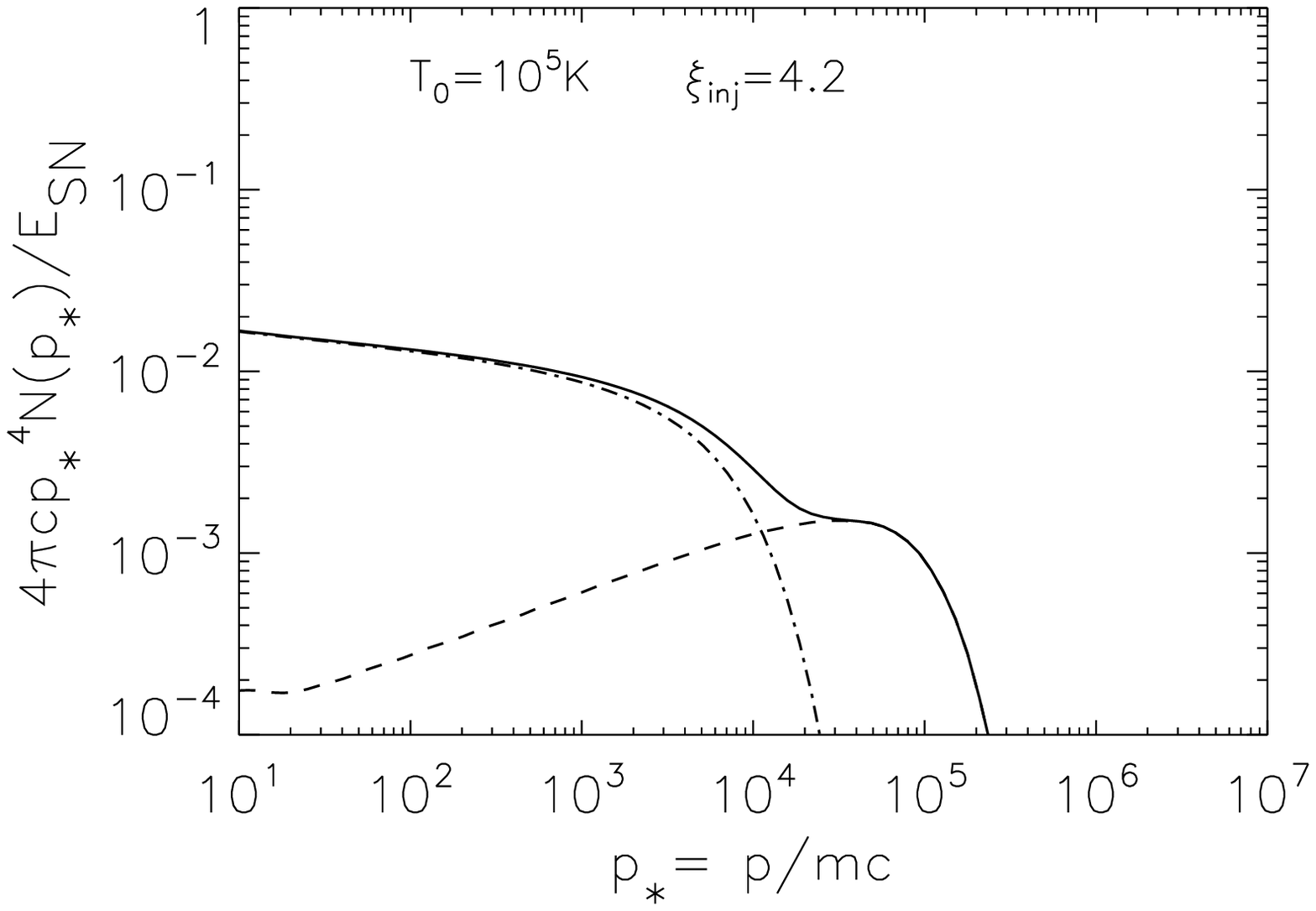}
\caption{Injection spectrum for a SNR exploding in a medium with density $n_0=0.1 \rm{cm}^{-3}$, temperature $T_{0}=10^{5}\dK$ and injection parameter $\xi_{inj}=3.6$ ({\it left panel}) and $\xi_{inj}=4.2$ ({\it right panel}). For both panels $x_{0}=R_{sh}$ and the lines are labelled as in Fig.~\ref{fig:T5xi39}.} 
\label{fig:T5x01}
\end{center}
\end{figure}

The way the spectrum of injected particles is affected by changing the injection efficiency is illustrated in Fig.~\ref{fig:T5x01}: the left panel refers to $\xi_{inj}=3.6$, corresponding to injecting into the acceleration process a fraction $\eta\sim2\times10^{-4}$ of the particles crossing the shock, while the right panel is obtained with $\xi_{inj}=4.2$ ($\eta\sim2\times10^{-6}$). The benchmark case $\xi_{inj}=3.9$ corresponds to $\eta\sim2\times10^{-5}$.
One can see that in the less efficient case ($\xi_{inj}=4.2$) the resulting spectrum is steeper, lower values of the maximum momentum are reached and the energy channelled into accelerated particles is very low. It is difficult to notice any appreciable changes in the spectral shape between the case $\xi_{inj}=3.6$ and those in Fig.~\ref{fig:T5xi39}, though the most efficient case ($\xi_{inj}=3.6$) leads to a slightly higher particle flux as could be expected.  
 
\begin{figure}
\begin{center}
\includegraphics[width=0.49\textwidth]{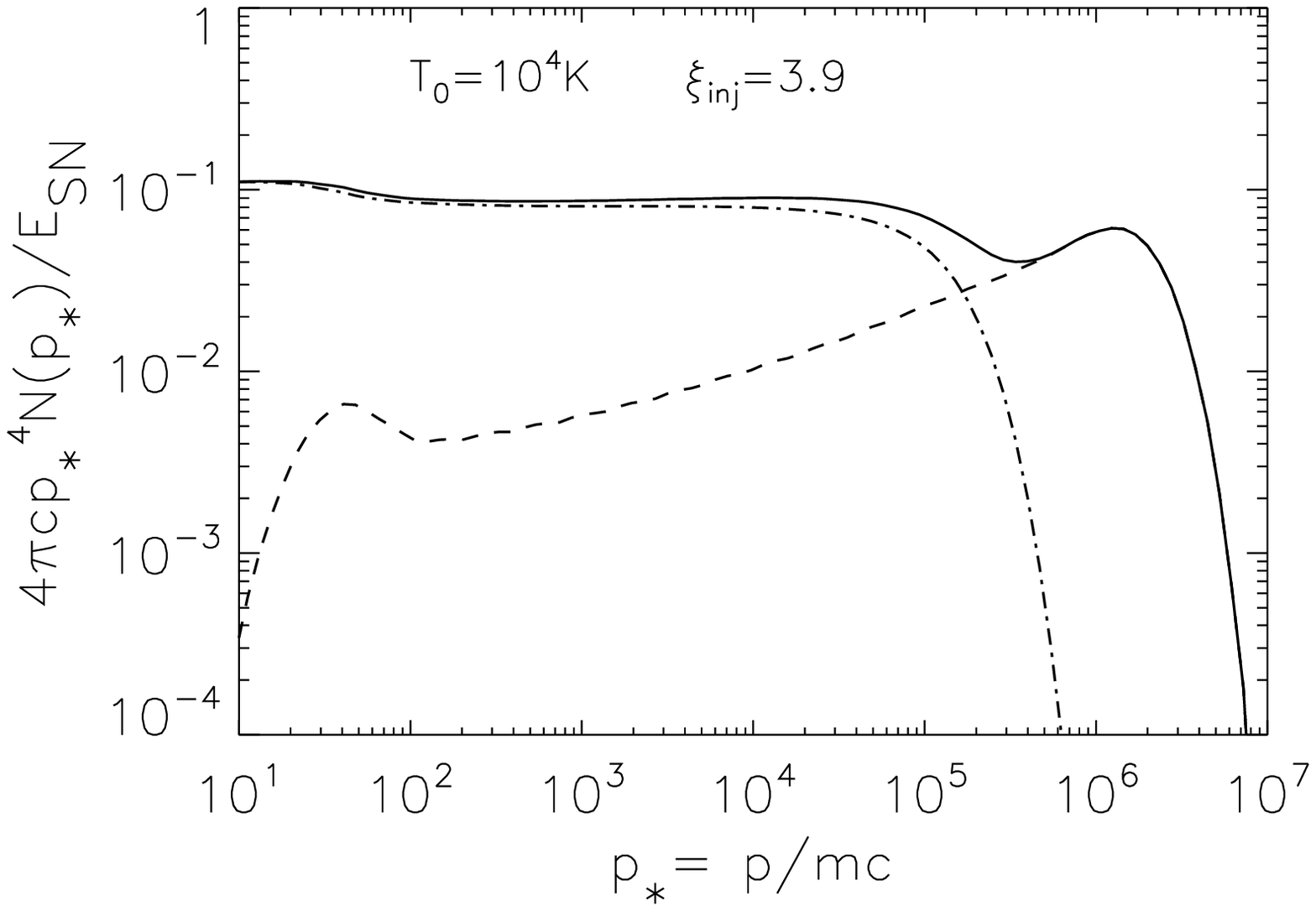}
\includegraphics[width=0.49\textwidth]{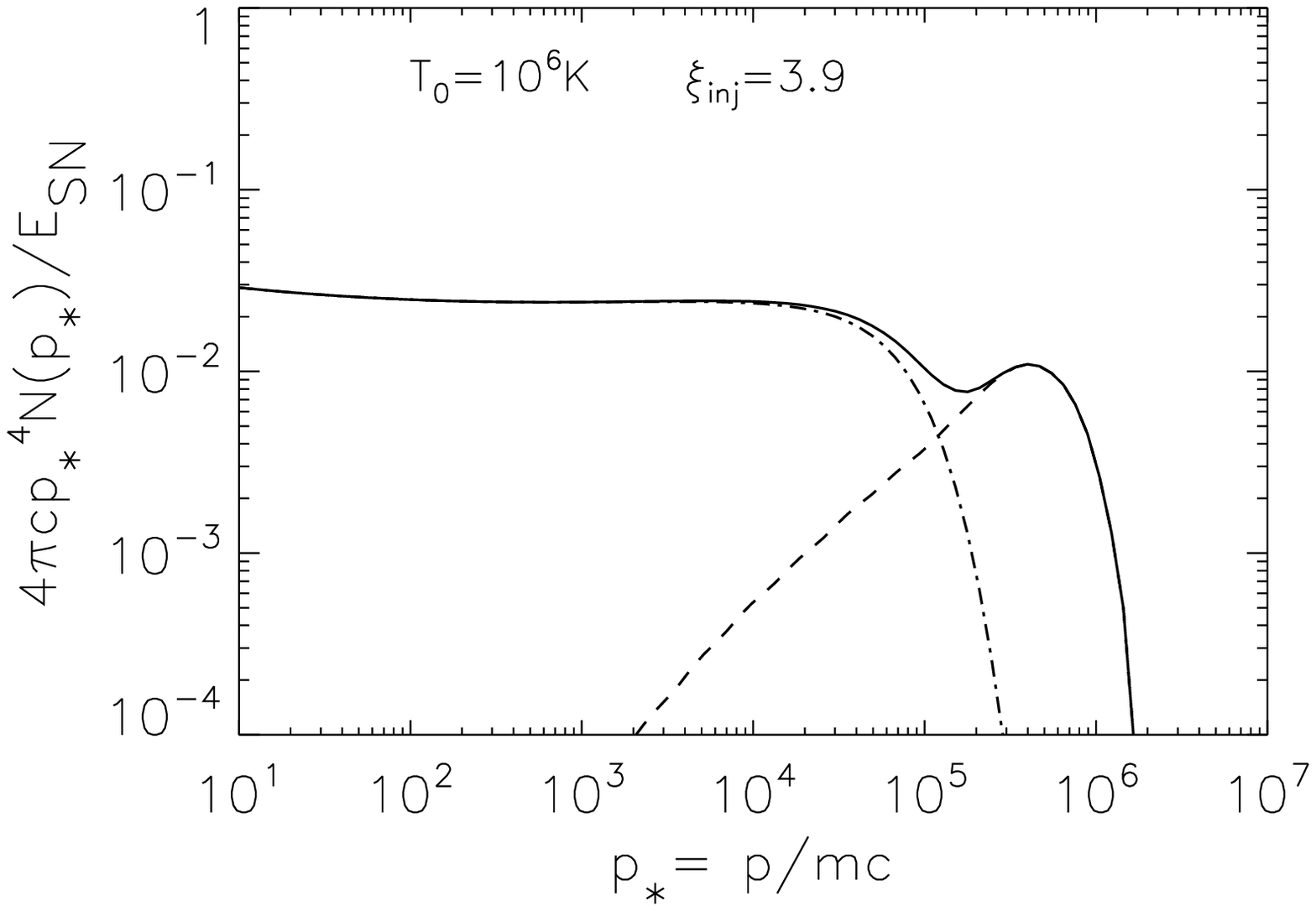}
\caption{Injection spectrum for a SNR exploding in a medium with temperature $T_{0}=10^{4}\dK$ (left panel) and $T_{0}=10^{6}\dK$ (right panel). In both cases the injection parameter is $\xi_{inj}=3.9$ and $x_{0}=R_{sh}$.} 
\label{fig:T4T6}
\end{center}
\end{figure}

As stressed in the initial discussion, we focus here on SNRs expanding in a spatially homogeneous medium, similar to the environment in which a type Ia SN is expected to occur. On the other hand it is interesting to explore the effects of warmer, more tenuous media on the particle acceleration process. In Fig.~\ref{fig:T4T6} we show the injected spectra for a medium with temperature $T_0=10^{4}\dK$ and gas density $n_0=1\rm{cm}^{-3}$ and one with $T_0=10^{6}\dK$ and $n_0=0.01\rm{cm}^{-3}$. Also these two cases show a total spectrum which is very close to a power law $p^{-4}$ up to $\sim 10^{5}$ GeV, with a bump close to the maximum energy reached during the SNR evolution. The shape of the escape flux from upstream is different in the two cases because of the very different values of the Mach number. In the case of a hot medium (right panel) the Mach number is systematically lower and not only the spectrum is somewhat steeper, but more important the escape flux drops faster when the Mach number drops in time. This leads to spectra of escaping particles which are more concentrated around the highest momenta. On the other hand the overall shape of the spectrum remains close to a power law although acceleration is somewhat more efficient (and the maximum momentum is higher) in the lower temperature case (left panel). 

The cases illustrated so far suggest that the spectrum injected by a SNR as a result of the integration over time of its injection history is very close to a power law $p^{-4}$ in the energy region where most high quality measurements are currently available. The good news is that the concavity which follows from the formation of a precursor upstream of the shock is not prominent in the injected spectra. The bad news is that it appears to be very difficult to steepen this injected spectra to the levels that are suggested by naive estimates based on simple diffusion models. We will discuss this point in \S \ref{sec:earth}.

\begin{figure}
\begin{center}
\includegraphics[width=0.8\textwidth]{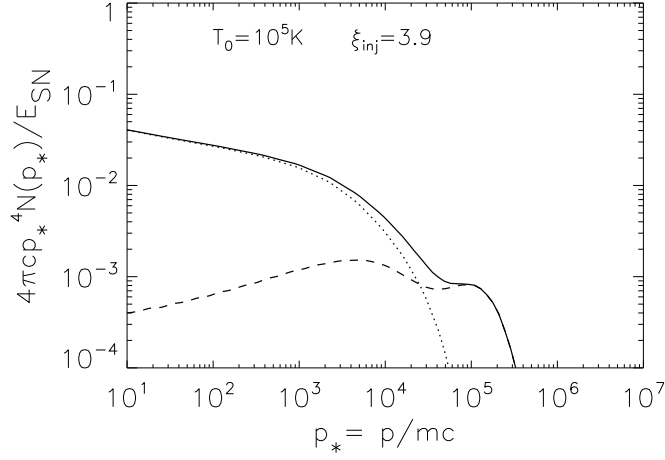}
\caption{Injected spectrum with forward moving waves downstream, with the Alfv\'en velocity v$_{A}$ calculated using the amplified magnetic field.} 
\label{fig:T5VA}
\end{center}
\end{figure}

An exception to the persistence of a very flat power law appears if one takes into account the finite speed of the waves responsible for particle scattering upstream and downstream. 
This point was discussed for instance by \cite{bell78} but it is easy to understand that the conclusions are very much model dependent. The spectrum of accelerated particles (even in the test particle theory of DSA) is determined by the compression factor of the velocities of the {\it scattering centers}. These centers are in fact plasma waves propagating in the upstream and downstream fluids and their velocity depends on the nature of the waves and on whether they are produced {\it in situ} and/or produced somewhere else and eventually advected. 
For instance the standard picture of NLDSA assumes that these waves are backward (i.e. moving against the fluid) Alfv\'en modes generated upstream of the shock and then partly reflected and partly transmitted through the shock surface, so that downstream there are only waves that have been advected from upstream \cite{sb71}. 
In this case one can show that the resulting effect on the spectrum of accelerated particles mainly consists in a flattening (e.g. \cite{schlick}). 
On the other hand, if gradients in the accelerated particles were present downstream, some level of turbulence could be generated downstream as well, so that there could be waves traveling away from the shock surface. 
In this scenario, and if the wave velocity is large enough, the spectrum of accelerated particles could be appreciably steeper, as investigated e.g. in \cite{zp08}. 
In Fig.~\ref{fig:T5VA} we show the injected spectrum in the case in which we assume that the waves downstream move in the forward direction at the Alfv\'en speed as calculated in the amplified field. 

A byproduct of dealing with steeper instantaneous spectra is that the shocks are less modified, the magnetic field amplification is less efficient and eventually the integrated spectrum is cut off at relatively low energy, $\sim 10^{4}-10^{5}$ GeV.

\subsection{Escape of particles at $p>p_{esc}(t)$}

Here we discuss the case in which at any given time all the particles with momentum $p>p_{esc}(t)$ escape the SNR, with $p_{esc}(t)$ defined so that $\lambda_{2}(p_{esc})\equiv D(p_{esc},B_{2})/V_{2}=x_{0}$, where $V_{2}=V_{sh}/R_{tot}$ is the downstream velocity.
The product $V\delta B$ is constant across the subshock, hence the diffusion length at any given $p$ immediately upstream of the shock is exactly the same as downstream.
On the other hand, the local diffusion length in the precursor $\lambda(x,p)\propto p/\delta B(x)/V(x)$ would be constant if only adiabatic compression were taken into account, but increases with distance from the shock as soon as CR-induced magnetic field amplification is included.
Since $p_{max}$ is determined by an average diffusion length throughout the precursor, the inequality $p_{esc}\geq p_{max}$ follows immediately. 
This implies that at any given time particles with momentum larger than a given ``escape'' momentum cannot be confined in the system.

\begin{figure}
\begin{center}
\includegraphics[width=0.49\textwidth]{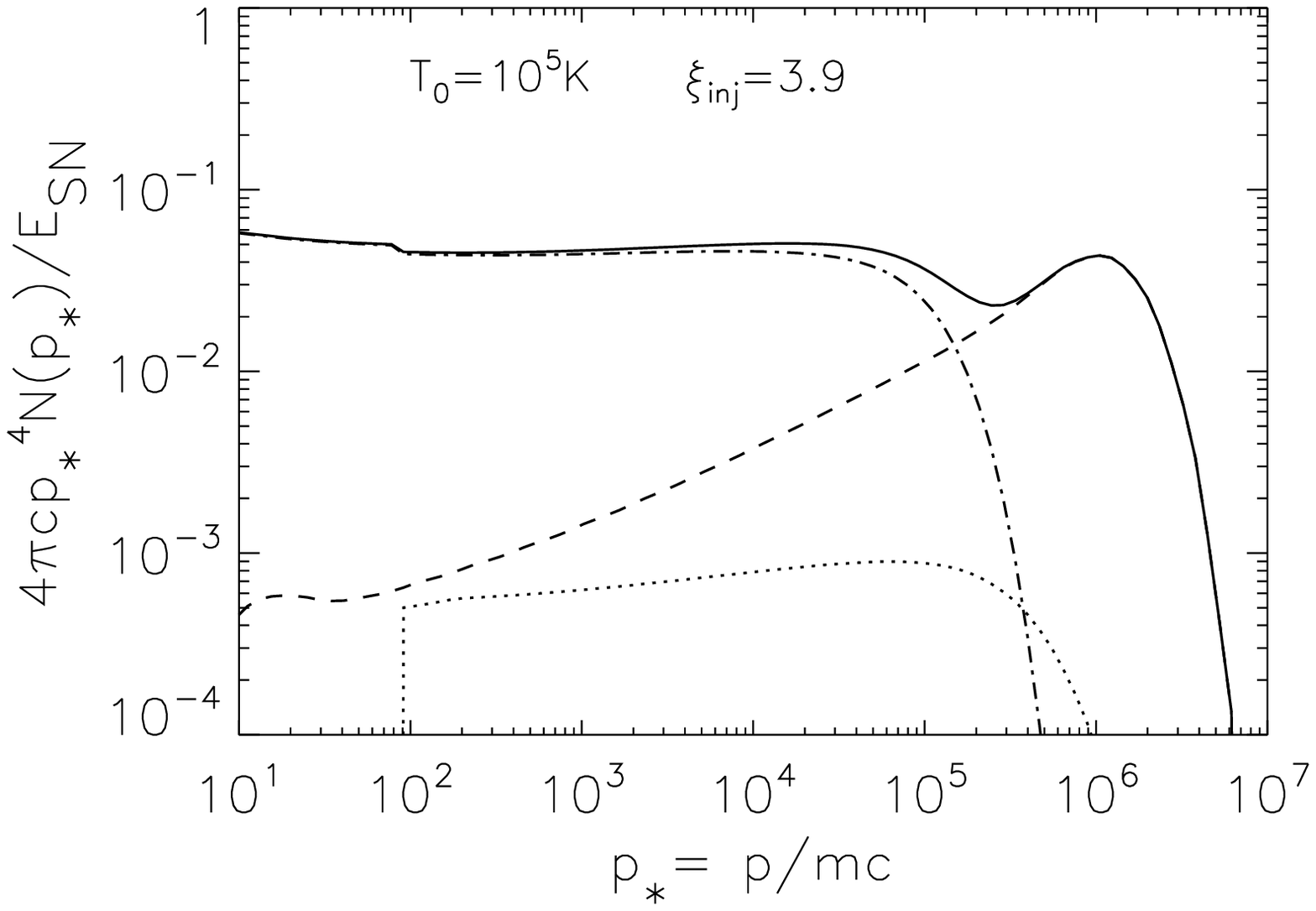}
\includegraphics[width=0.49\textwidth]{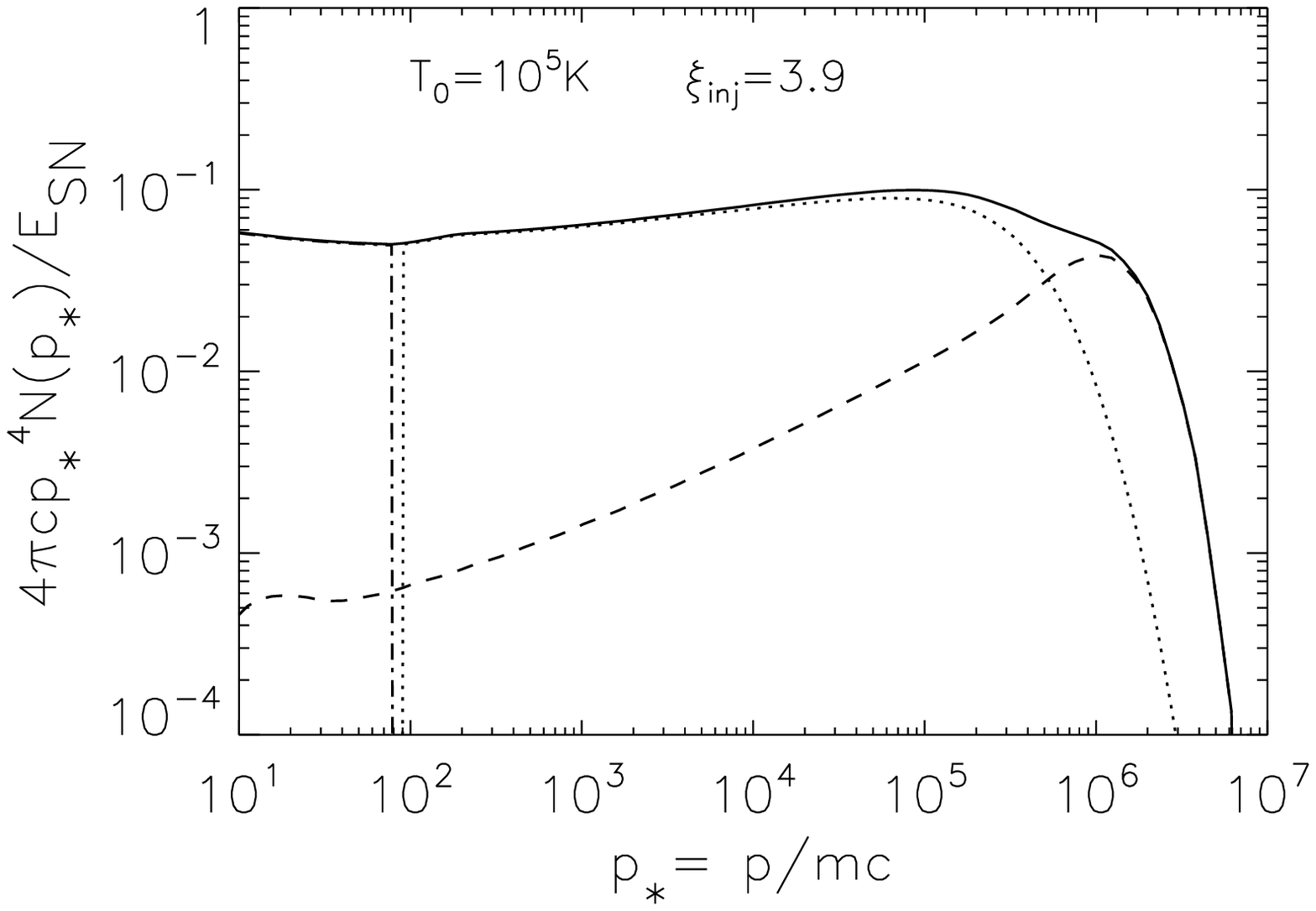}
\caption{Spectrum injected by our benchmark SNR if a fraction of particles with $p>p_{esc}(t)$ (all particles in the right panel and 10\% of them in the left panel) leave the accelerator at any given time. In both panels the dashed line represents the escape flux through the free escape boundary, the dotted line is the flux of particles escaping at $p>p_{esc}(t)$ and the dash-dotted line refers to the particles that remain in the expanding shell and escape at the end of the evolution. The solid line is the sum of all contributions. 
} 
\label{fig:escall}
\end{center}
\end{figure}

In Fig.~\ref{fig:escall} we show the spectrum injected by an individual SNR in this scenario. We assume that the SN explosion occurs in a medium with magnetic field $B_{0}=5\mu$G, temperature $T_0=10^{5}\dK$ and density $n_0=0.1 \rm{cm}^{-3}$ and the injection parameter is $\xi_{inj}=3.9$: these are the benchmark parameters already used to obtain the results shown in Fig.~\ref{fig:hydro} and Fig.~\ref{fig:spectra}. The free escape boundary is assumed to be at $x_0=R_{sh}$. The two panels of Fig.~\ref{fig:escall} refer to the case in which all particles with $p>p_{esc}(t)$ escape the accelerator at any given time (right panel) and to the case in which only 10\% of them are allowed to escape the acceleration region (left panel). In this latter case, the particles that are trapped in the shell are advected downstream and lose energy adiabatically. In both panels the dashed line represents the escape flux through the free escape boundary, the dotted line is the flux of particles escaping at $p>p_{esc}(t)$ and the dash-dotted line refers to the particles that remain in the expanding shell and escape at the end of the evolution. The solid line is the sum of all contributions. 

It is clearly visible that the net effect of the instantaneous escape at $p>p_{esc}$ is to flatten the injected spectrum and possibly wash out the dip-like feature at the highest energies (see right panel). These findings seem to agree with the results of previous calculations presented in \cite{pz05}, where a similar recipe for escape was adopted.

Our conclusion is that even in this escape scenario the generic spectrum of injected particles is very close to $p^{-4}$ or flatter. 

\subsection{Escape from a broken shell}

\begin{figure}
\begin{center}
\includegraphics[width=0.49\textwidth]{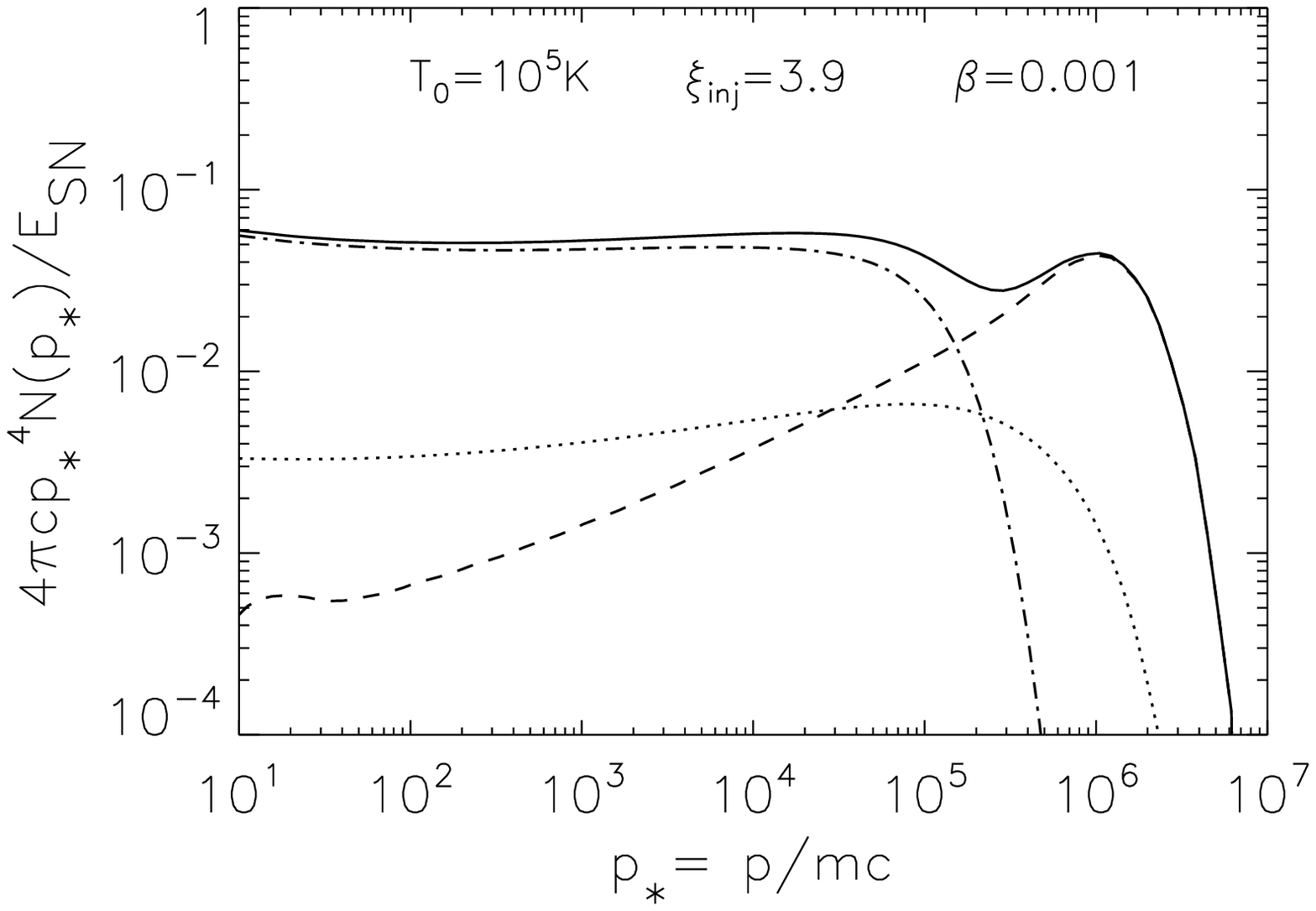}
\includegraphics[width=0.49\textwidth]{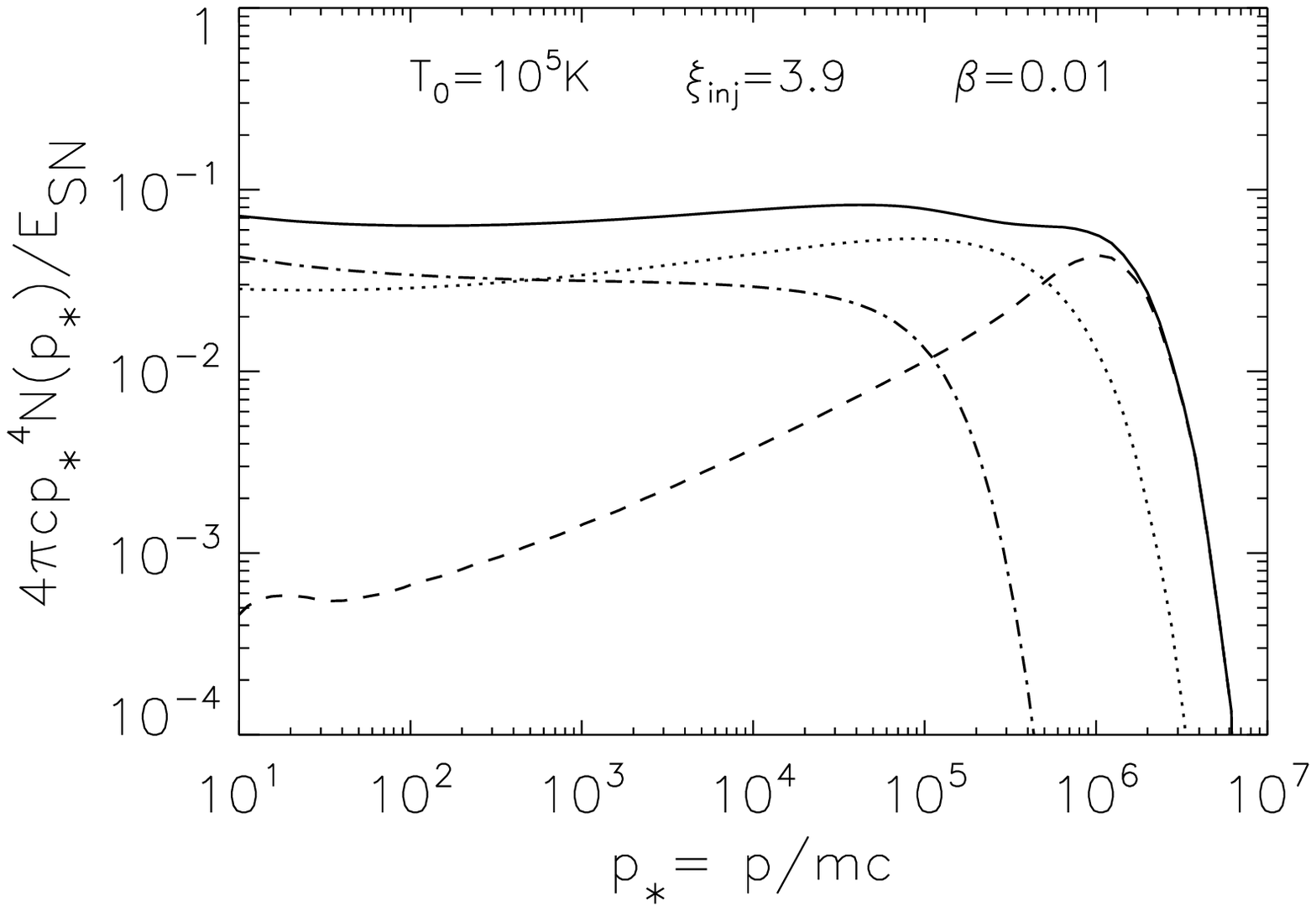}
\includegraphics[width=0.49\textwidth]{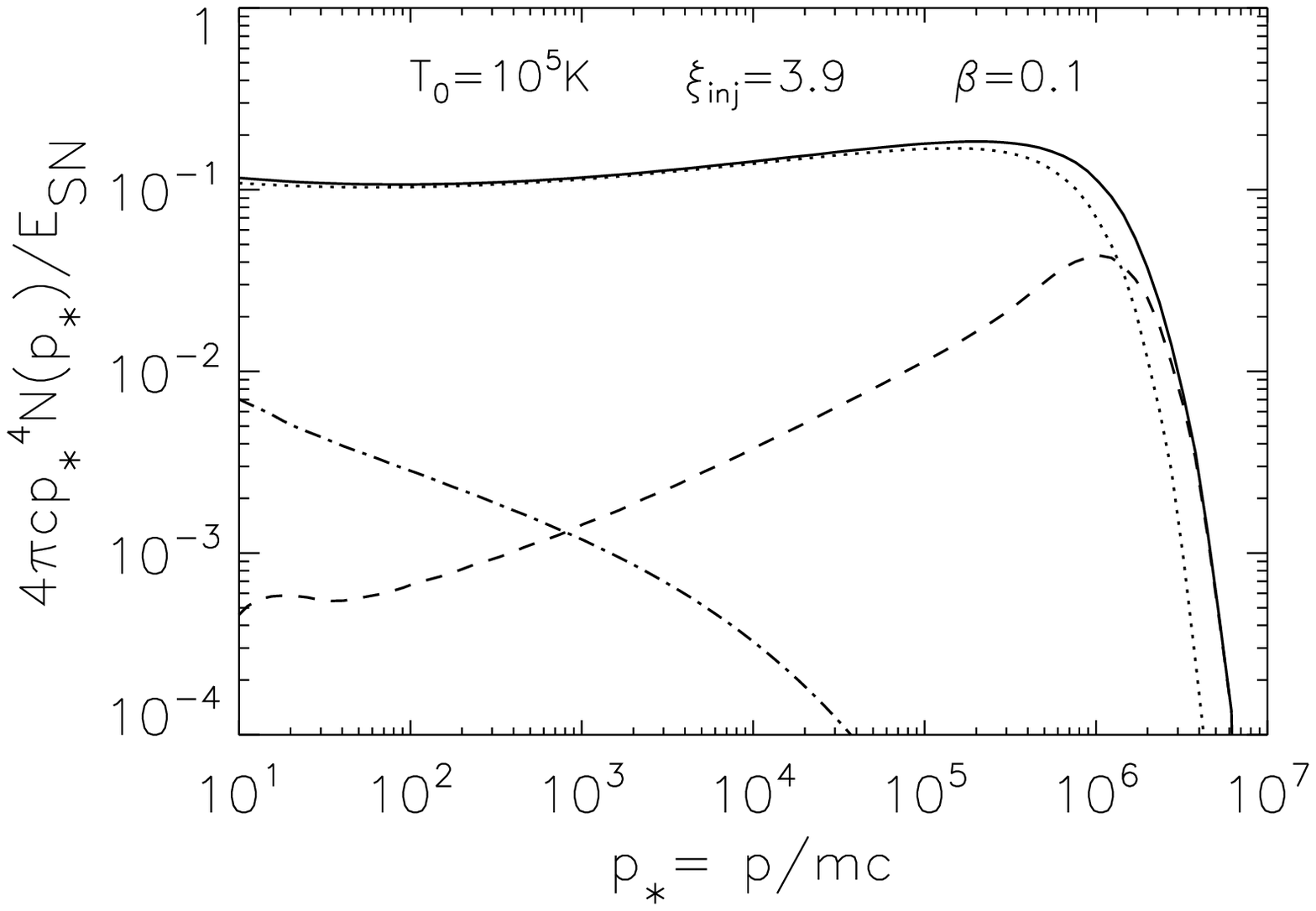}
\caption{Spectrum injected by our benchmark SNR if particles escape the acceleration region from a broken shell. The three panels refer to four values of $\beta$ as indicated, where $\beta$ is the fraction of particles that escape the shell from downstream. In all panels the dashed line represents the escape flux through the free escape boundary, the dotted line is the flux of particles escaping from the sides and the dash-dotted line refers to the particles that remain in the expanding shell and escape at the end of the evolution. The solid line is the sum of all contributions. 
} 
\label{fig:broken}
\end{center}
\end{figure}

Here we discuss the case in which the expanding shell is broken, possibly due to instabilities and/or inhomogeneities in the circumstellar medium. In this case, at any given time particles can escape the expanding shell from the sides in addition to escaping from the far upstream region.

In Fig.~\ref{fig:broken} we show the spectrum injected by an individual SNR in this scenario. We consider again the benchmark environmental parameters: $B_{0}=5\mu$G, $T_0=10^{5}\dK$, $n_0=0.1 \rm{cm}^{-3}$, $x_0=R_{sh}$ and $\xi_{inj}=3.9$. The three panels refer to different values of the parameter $\beta$ (as indicated) which quantifies the fraction of the particles in the downstream plasma that are allowed to escape the system. For $\beta<1$, the particles that are unable to escape are advected downstream, lose energy adiabatically and are injected at the end of the SNR evolution as usual. 
As one could easily expect, while $\beta$ increases the gap between the escape flux to upstream infinity (dashed lines) and the advected spectrum is filled and eventually disappears for $\beta=0.1$. 
When this happens, however, the time integrated spectrum injected by the SNR is visibly flatter that $p^{-4}$.

\section{The spectrum at Earth}
\label{sec:earth}

The spectrum of CRs observed at Earth is the result of several complex phenomena occurring during propagation: particles are injected at the sources, for instance in SNRs, then diffuse in the interstellar magnetic field and could possibly be advected in a Galactic wind if one is present \citep{jones79}. 
Moreover CRs may be reaccelerated during propagation due to second order Fermi acceleration induced by scattering against Alfv\'en waves in the Galactic magnetic field \citep[e.g.][]{seoptu}. In principle all these phenomena modify the spectrum with respect to the injected one. For standard values of the parameters, both advection in a wind and reacceleration become of some importance only at relatively low energies and for the purpose of the present discussion they can be safely disregarded \citep[see e.g.][and references therein]{jones+01}. 

If the diffusion coefficient has the form $D(E)\propto E^{\delta}$, the spectrum observed at Earth can be estimated as being $N(E)\propto Q(E) E^{\delta}$, so that for an injection spectrum $Q(E)\propto E^{-2}$ the observed spectrum requires $\delta\sim 0.65-0.7$. 

This simple estimate leads to an important consequence, recently reviewed in \cite{hillas}, that CRs at Earth should become highly anisotropic at energies much lower than the knee. This anisotropy is not observed, hence posing a very serious problem for simple recipes of CR diffusion in the Galactic magnetic field. In order to alleviate this problem it has been proposed that the injection spectrum could be $Q(E)\propto E^{-2.4}$ and that the diffusion coefficient could be $D(E)\propto E^{1/3}$, but as we discussed above, at least in the case of SNRs or any other source where the acceleration process is based on the first order Fermi process in highly supersonic shocks, this situation is very hard to reproduce since the derived injection spectra are generically harder than $E^{-2.4}$. 

On the other hand one should keep in mind that the conclusion on the anisotropy might be due to too simplistic leaky box models or diffusion models \citep[e.g.\ GALPROP][]{galprop} where the Galactic magnetic field has no structure: the interplay between parallel and perpendicular diffusion and the random walk of magnetic field lines could for instance have a crucial influence on the anisotropy of CRs. Moreover, as discussed in \cite{ptuskin2006}, the observed anisotropy could be affected in a non trivial way by the distribution in space and time of local supernovae. 

The rather disappointing picture that arises from this line of thought is that, even if the basic principles of both acceleration and propagation of CRs are thought to be rather well understood, at the present time neither the injected spectrum nor the propagated spectrum can be reliably calculated. The main obstacle to reaching clear predictions is in the complex nature of the accelerator and of the Galaxy as a medium in which CRs propagate. Progress on the first issue is likely to come from efforts aimed at clarifying the nature of the turbulent magnetic field that is responsible for particle scattering and escape: 1) theoretical investigation is required of the instabilities that are more likely to lead to amplified magnetic fields at scales that are useful for the particle scattering; 2) precious information is still to be gathered from comparison between observations and models of the multifrequency emission of individual sources \citep[e.g.][]{mor09}, including morphological information \citep[e.g.][]{mor10}).    

\section{Discussion}

Here we discuss the main ingredients and uncertainties that enter the calculation of the spectrum of cosmic rays injected by SNRs. This is determined by the superposition of two contribution: 1) particles that escape the expanding shell from a free escape boundary at some location upstream of the shock; 2) particles that leave the accelerator at some later time when the shock slows down and liberates the particles trapped behind it. 
The latter phenomenon takes place only after the shell has expanded and particles behind the shock have suffered adiabatic energy losses. This is a crucial point because if indeed SNRs, at some stage of their evolution, are able to accelerate CRs (protons) up to the knee energy, these particles cannot contribute to the CR spectrum around the knee unless they leave the accelerator immediately after production. This short introduction already opens the way to several questions: 1) where is the free escape boundary located? 2) what physical processes regulate its position? 3) which particles escape the accelerator at any given time? 

As illustrated by the numerous cases considered in this paper, although we have phenomenological tools to calculate what might happen, we are not able at the present time to provide unique answers to the questions above. 

Let us start the discussion with a comment on the commonly adopted recipe to describe the escape of particles by assuming the existence of a free escape boundary at some location $x_{0}$ upstream of the shock. While from the mathematical point of view, this assumption is well posed,
from the physical point of view the problem remains in that the position of this boundary is related to poorly understood details of the problem, especially the ability of particles to self-generate their own scattering centers. 
The position of the free escape boundary should in principle coincide with a location upstream of the shock where particles are no longer able to scatter effectively and return to the shock. This would lead to an anisotropic distribution function of the accelerated particles, that can no longer be described by the standard diffusion-convection equation. Moreover, while waves can be generated both resonantly \cite{skillinga,bell78} and non-resonantly \cite{bell04}, particles can scatter effectively only with resonant waves. This adds to the complexity of the problem, in that one might have amplified magnetic fields of large strength but on scales which do not imply effective scattering of the highest energy particles. This concept of a free escape boundary which is self-adjusted by the accelerated particles adds to the extreme non-linearity of NLDSA and is currently not included in any of the calculations presented in the literature. This clearly makes the prediction of a maximum energy of accelerated particles very uncertain whenever it is determined by the size of the accelerator (namely by $x_{0}$) rather than by the finite age of the accelerator. 

What appears to be a rather solid result is that the highest maximum energy throughout the history of the SNR is reached at the beginning of the Sedov-Taylor phase, provided the magnetic field is self-generated by the accelerated particles through streaming instability. However the nature of the mechanism responsible for the magnetic field amplification is unknown: the bright narrow X-ray rims suggest that the interstellar magnetic field is amplified at the shock, but at the present time it is not possible to say for sure whether the field is induced by CRs or by some type of fluid instability associated with the corrugation of the shock surface due to the propagation in an inhomogeneous environment (see for instance \cite{joki07}). On the other hand, even if the magnetic field is induced by the presence of accelerated particles, the flavor of CR induced instability involved is all but trivial to identify. Resonant streaming instability, the only one included in the calculations presented here, has the advantage of producing waves which are at the right wavelengths to scatter particles resonantly, thereby increasing their energy because of multiple shock crossings. However particles can reach the energy of the knee only if the mechanism is assumed to work efficiently even in the regime in which the field has reached non-linear amplification, $\delta B/B_{0}\gg 1$, which is all but obvious since the resonance condition becomes ill defined in this regime. 

Non-resonant magnetic field amplification (e.g. \cite{bell04}) can possibly lead to larger values of the turbulent magnetic field, but typically the field is produced on scales which are minuscule compared with the gyration radius of the highest energy particles, which makes it hard to understand how they reached that energy in the first place and how they can keep increasing their energy, unless a very effective inverse cascade occurs in the precursor, thereby transferring power to larger spatial scales.

The general trend of the injection spectra calculated in this paper is to be very close to power laws with index -4, with all the difficulties that this implies in terms of connecting SNRs with CRs observed at Earth. On the other hand it is remarkable that quasi-power-law spectra are obtained by overlapping instantaneous spectra which are characterized by the concavity typical of NLDSA. This important physical point should be kept in mind whenever one tries to infer the spectrum of accelerated particles from that of the radiation observed by a SNR. In general the two spectra are not required to be the same.

A noticeable exception to the rule of injected spectra that are flat power laws is represented by the case in which the waves responsible for the scattering of accelerated particles in the downstream plasma move in the forward direction. This scenario would lead to a time integrated spectrum which is appreciably steeper than $p^{-4}$. However the instantaneous spectra are also rather steep, which implies that magnetic field amplification is not very efficient and the maximum momentum of accelerated particles is much lower than the knee (see Fig.~\ref{fig:T5VA}). Although the basic physical intuition associated with having a large velocity of scattering waves downstream is to infer that the spectra can become appreciably steeper (or flatter for that matter, it all depends on the direction of motion of the waves) one should also keep in mind that when $\delta B/B\gg 1$ and the waves are not necessarily Alfv\'en waves, even the form of the transport equation as is usually used might be profoundly affected: particles might propagate in a non diffusive way in the shock proximity. 

Another case in which we obtained relatively steep spectra injected by a SNR is that of injection with low efficiency (see the case $\xi_{inj}=4.2$ in Fig.~\ref{fig:T5x01}). However this case also leads to a rather small fraction of energy channelled into accelerated particles and to a rather low maximum momentum, which makes this case appear of scarce interest for the origin of CRs, at least in the context of the standard picture of CR propagation in the Galaxy.

A caveat for this type of calculation of the injection history of CRs in SNRs is that the surrounding medium could be much more complicated than assumed here. For instance in a type II SN one might expect that the shell propagates first in the magnetized wind of the presupernova star where the magnetic field should be mainly perpendicular. In this case particle acceleration does not occur in the regime described by the transport equation used here (and in most literature on the topic). Drifts in the shock region might make the maximum energy achievable higher than predicted by NLDSA at parallel shocks (see for instance \cite{joki87}). At some time in the evolution one could envision a transition to a mainly parallel field configuration where our calculations would apply. The time integrated spectrum in this case could be different from those calculated here which apply to standard type Ia SN. The case of type II SN has been mimicked here only by assuming a hot, more rarefied circumstellar medium. 

On the other hand, it is conceivable that a spread in the acceleration efficiencies (and/or in the maximum achievable momenta) between individual sources might be essential to explain the overall observed spectrum of Galactic CRs. If this is the case, the work presented in this paper is only a first step towards reproducing the observations, a task that can only be accomplished by adding up the contributions due to different populations of SNRs, with different environmental parameters (see e.g. in Fig.~5 of \cite{hillas}).

\section*{Acknowledgments}
This work was partially supported by MIUR (under grant PRIN-2006) and by ASI through contract ASI-INAF I/088/06/0. This research was also supported in part by the National Science Foundation under Grant No. PHY05-51164. We wish to acknowledge the KITP in Santa Barbara for the exciting atmosphere during the Program {\it Particle Acceleration in Astrophysical Plasmas}, July 26-October 3, 2009.

\end{document}